# Evaluating Integrative Strategies for Incorporating Phenotypic Features in Spatial Transcriptomics

Levin M Moser*[1,2], Ahmad Kamal Hamid*[1], Esteban Miglietta[1], Nodar Gogoberidze[1], Beth A Cimini[1+]

1=Imaging Platform, Broad Institute of MIT and Harvard, Cambridge MA, USA
2=ETH Zurich, Zurich, Switzerland
*= equal contribution
[+]= to whom correspondence should be addressed; Contact details:

Dr. Beth Cimini,
Imaging Platform,
Broad Institute,
415 Main St,
Cambridge, MA 02142
Email: bcimini@broadinstitute.org
Phone: 617-714-7000

## ABSTRACT

The key advantage of spatial transcriptomics (ST) technologies lies in the spatial domain: these techniques not only offer an unprecedented opportunity to interrogate intact biological samples in a spatially informed manner, but also set the stage for integration with other imaging-based modalities. However, how to most effectively exploit spatial context and integrate ST with imaging-based modalities that capture morphological insight remains an open and heavily investigated question. To address this, particularly under real-world experimental constraints such as limited dataset size, class imbalance, and bounding-box-based segmentation, we used a publicly available murine ileum Multiplexed Error-Robust Fluorescence In Situ Hybridization (MERFISH) dataset to evaluate whether a minimally tuned variational autoencoder (VAE) could extract informative low-dimensional representations from cell crops of spot counts, nuclear stain, membrane stain, or a combination thereof. We assessed the resulting embeddings through PERMANOVA, cross-validated classification, and unsupervised Leiden clustering, and compared them to classical image-based feature vectors extracted via CellProfiler. While transcript counts (TC) generally outperformed other feature spaces, the VAE-derived latent spaces (LSs) captured meaningful biological variation and enabled improved label recovery for specific cell types. LS2, in particular, trained solely on morphological input, also exhibited moderate predictive power for a handful of genes in a ridge regression model. Notably, combining TC with LSs through multiplex clustering led to consistent gains in cluster homogeneity, a trend that also held when augmenting only subsets of TC with the stain-derived LS2. In contrast, CellProfiler-derived features failed to match the performance of the LSs, highlighting the advantage of learned representations over hand-crafted features. Collectively, these findings demonstrate that even under constrained conditions, VAEs can extract biologically meaningful signals from imaging data and constitute a promising strategy for multi-modal integration.

## LAY DESCRIPTION

Spatial transcriptomics (ST) encompass technologies that measure gene expression in tissue samples without disrupting their organization, unlike other single-cell techniques. In imaging-based ST specifically, images of cell nuclei and boundaries are often produced in the process, providing insight into both cellular spatial relationships and morphology (shape, texture, etc.) while also being aligned with the high-resolution spatially resolved gene expression data. Although these modalities are inherently compatible and offer complementary information that may, collectively, improve biological insight and tasks like cell type identification, there is currently no universally established method for integrating them. Here, we explored using a deep learning model for this integration task and compared its performance to classical image feature extraction methods, focusing on real-world constraints such as small datasets with diverse and disproportionately abundant cell types, and stain images where cells are cropped to their approximated bounding box as opposed to their precise membrane borders. To that end, we trained a variational autoencoder (VAE)—a neural network that learns compact representations of its input—on various combinations of spatial gene expression data, nuclear stains, and cell border stains from a publicly available dataset generated from a single mouse intestinal sample. As the dataset also included ground truth cell identities, we assessed how well the VAEs captured biologically relevant features through statistical testing and supervised classification, comparing the learned representations to both the original gene expression data and classical image features generated with CellProfiler. Despite the constraints, the VAEs captured substantial biological variation, outperforming classical feature extraction with representations of stain images alone being able to predict the expression of some genes. Finally, we investigated whether these feature spaces could distinguish cell types in an unsupervised setting, a common goal in spatial biology studies. Multiplex clustering showed that combining gene expression data with VAE representations improved the purity of cell type groupings compared to using gene expression alone. These findings suggest that even relatively simple deep learning models can extract meaningful biological information from imaging data and offer a promising strategy for integrating imaging and gene expression in spatial transcriptomics.

# KEYWORDS

Spatial transcriptomics, multi-modal integration, variational autoencoder, cell type deconvolution, morphological features

# INTRODUCTION

In recent decades, biological and medical research have witnessed remarkable advancement, propelled largely by the development of increasingly powerful measurement technologies. From omics platforms to advanced imaging, such highly resolved and highly dimensional methods have dramatically transformed the interrogation of cellular and tissue-level processes, advancing the understanding and treatment of complex diseases such as cancer and Alzheimer's disease [1,2]. Born from the recognition that spatial and structural context is critical for interpreting biological systems, spatial transcriptomic (ST) methods have gained widespread use, with remarkable success [3–5]. By preserving tissue architecture and capturing gene expression *in situ*, ST additionally avoids artifacts introduced by tissue dissociation, offering a more faithful view of cellular organization and function [6]. These advantages are further amplified in imaging-based ST, which enable the pairing of ST data with high-resolution morphological context [7]. However, despite the clear promise of ST technologies, effective multi-modal integration of transcriptional, spatial, and morphological information remains an open challenge, as disparities in statistical distributions, feature count, and scale continue to hinder appropriate data harmonization.

To surmount the integration challenges posed by ST datasets, a growing number of studies have turned to deep learning (DL) approaches, which are inherently well-suited to robustly handle complex, high-dimensional, and heterogeneous data types [8]. While showcasing compelling success, these studies mainly focused on integrating ST with histological stains (e.g. ConGI [9], TESLA [10], stLearn [11], TransformerST [12], and Starfysh [13]), chromatin images (e.g. STACI [14]), scRNA-seq (e.g. ENVI [15], STEM [16], GTAD [17], and GraphST [18]), or multi-omic datasets (e.g. PRAGA [19] and SpatialGlue [20]) towards tasks such as spatial domain identification, gene imputation, cross-modal prediction, biomarker discovery, or cell type deconvolution. On the other hand, there is a notable paucity in studies investigating the multi-modal integration of ST with paired nuclear and membrane stain images, particularly in the context of real-world experimental conditions, where the objective is often cell type deconvolution and classification despite constraints such as limited dataset size [5], skewed cell type abundance [21], and approximated cell segmentation [22].

Historically, cell type identification and classification have evolved considerably: from qualitative assessments of histological morphology, to the use of protein markers via immunofluorescence, and, more recently, to the quantitative profiling of single-cell transcriptomes. However, despite these advancements, transcriptomic data does not diminish the relevance of morphological information as the two modalities capture distinct and complementary aspects of cell state and identity [23]. In this study, we aimed to explore how these complementary modalities can be jointly leveraged to improve cell type deconvolution under real-world constraints, laying the groundwork for more accurate cell state classification in studies comparing healthy and diseased tissue. Specifically, we trained several variational autoencoders (VAEs) on a publicly available Multiplexed Error-Robust fluorescence *in situ* hybridization (MERFISH) dataset derived from a single murine ileal tissue sample [24,25], using paired spot count, nuclear stain, and membrane stain images, or a combination thereof. We then studied the potential of the resulting latent spaces (LSs)—low-dimensional latent representations—to inform cell type deconvolution using various approaches including joint unsupervised clustering with transcript counts (TC), against the published ground truth (GT) annotations. Moreover, to contextualize our findings, we compared VAE-derived features to classical features extracted with CellProfiler (CP) [26,27], a well-established open-source tool for robust and unbiased phenotypic profiling across diverse image-based datasets [28,29]. Overall, our findings revealed context-specific strengths of VAE-derived morphological embeddings in constrained settings, showing that even minimal imaging data can meaningfully augment ST datasets.

## METHODS

### Dataset and pre-processing

All experiments were conducted using a publicly available MERFISH dataset generated from a single murine ileal tissue sample [24,25]. The dataset encompasses spot counts for 241 genes, paired nuclear (DAPI) and membrane stain images (immuno-labelled Na$^+$/K$^+$-ATPase with a sequence-tagged secondary antibody), and a GT partition of 19 cell types, assigned using the CellAnnotatoR package and a manually assembled marker list. Segmentation and preprocessing were performed in the original study and are included in the public dataset. Briefly, the Baysor algorithm was applied with the membrane stain as a prior, where nuclei were first segmented using CellPose and transcript spots were assigned to cells in a Bayesian probabilistic framework that optimized the joint likelihood of the transcripts and available auxiliary stains [30]. The convex hull of the spots assigned to each cell was then used to estimate its boundaries, and cells smaller than 50 pixels or with fewer than 10 assigned transcripts were removed from downstream analysis. Due to the lack of exact cell boundaries in Baysor segmentation, cell bounding-box crops were then generated from the stain images based on convex hull centroids.

In this study, unannotated cells in the dataset were excluded from further analysis, resulting in a final count of 5192 cells and a TC matrix sparsity of 89.62%. Using the Scanpy library [31], TC were first normalized such that total count per cell matched the median total count across all cells, and subsequently log1p-transformed. Various crop sizes were tested to assess the impact of spatial neighborhood information on VAE training and subsequent clustering analyses. For experiments where gene subsets were used, three selection approaches were adopted: 1) The principal component analysis (PCA)-based approach involved the identification of the top 3 genes in terms of absolute loadings for the first 15 PCs, selected based on the elbow point in the scree plot, resulting in 25 unique genes. 2) The GT-based approach involved the identification of the top gene per true label based on differential expression analysis with the Wilcoxon rank sum test, yielding 19 genes. 3) A random selection approach was also employed, with 19 randomly chosen genes.

## Models

*Learning framework*

Unlike most previous studies employing VAEs to learn latent embeddings of ST data [14,32,33], our approach incorporates the local spatial distribution of transcripts in the input rather than merely aggregating TCs into per-cell gene count vectors. While this introduces a more complex learning problem, retaining spatial information within each cell may preserve biologically relevant signals otherwise lost by gene count aggregation. As such, the VAE reconstructs decoded single-pixel MERFISH spots, a task complicated by inherent data sparsity, where small reconstruction errors lead to pixel misclassification and complete loss. To alleviate this, each spot was expanded to neighboring pixels, approximating transcript areas rather than discrete coordinates.

Two data representation strategies were explored to facilitate this learning paradigm. One strategy entailed encoding the expanded spots with their corresponding gene index in the MERFISH codebook effectively framing the task as a regression problem. However, due to poor reconstruction and embedding validation (data not shown), an alternative approach was adopted, reformulating the task as a multi-class semantic segmentation problem: cells were represented as multi-plane data objects, with each plane corresponding to a gene and containing a binary matrix indicating the presence (foreground) or absence (background) of expanded decoded MERFISH spots. This approach introduced two main challenges: a linear increase in computational burden and a further exacerbation of the sparsity problem. The former was deemed acceptable given the dataset size and the latter was mitigated through careful loss function design and corresponding class weighting, as detailed below.

All models were trained on the full dataset with no held-out validation set, effectively operating in a transductive learning framework. This deliberate decision was motivated by a prioritization of feature extraction over generalization: model overfitting is of minimal concern, as the dimensionality of the VAE bottleneck layer can be controlled to constrain LS capacity, preventing memorization while enforcing efficient learning. Additionally, since the LS is used for downstream analysis, the model's reconstruction ability is irrelevant to the aim of this study. In order to explicitly preclude the learning of positional features, a data augmentation strategy was imposed whereby both input TC objects and stain images underwent random orthogonal rotation.

*VAE architecture*

The convolutional VAE implemented in this study largely follows a standard architecture [34] (Figure 1): It consists of five convolutional blocks in both the encoder and decoder. Each encoder block includes a 2D convolutional layer with a kernel size of 3, a stride of 2, and no padding, followed by a leaky ReLU activation and batch normalization. This structure is mirrored in the decoder, with transposed convolutional layers to facilitate upsampling. The multi-modal VAE which combines the two data modalities (VAE3–5) is implemented following a double-decoder architecture: Both data modalities are mapped to the same latent space using a single encoder (using the same structure as outlined above). Reconstruction is done by using two separately trained decoders mapping the latent vector back to the respective data modalities. Furthermore, genes with fewer than 100 transcripts across the tissue were discarded in the multi-modal models (VAE3–5) to curb the model's complexity. While all the models were trained using a ReLU slope of 0.1 with the Adam optimizer [35] and a 50-dimensional LS, other architectural parameters and training hyperparameters were empirically adjusted but not systematically optimized (Table 1). Empirical experiments with different hyperparameters showed no observable effect in terms of embedding performance when tuning hyperparameters such as the batch size, learning rate, or optimization algorithm.

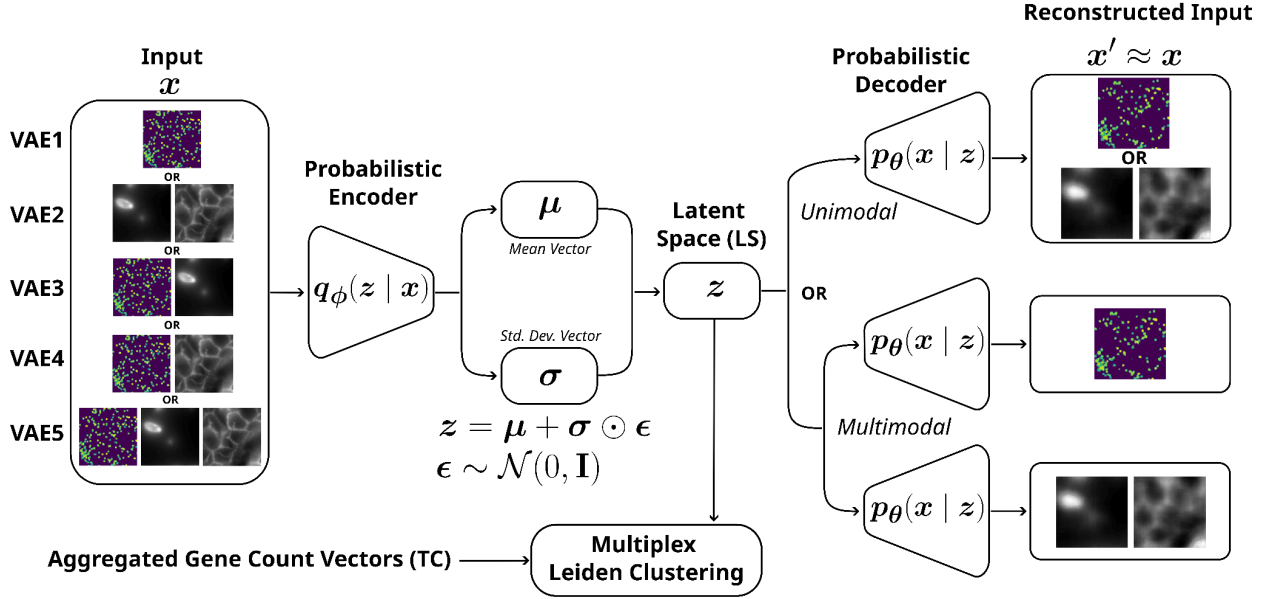

**FIGURE 1 | Abstraction of the VAE architecture showing the encoder-decoder framework and sampling process.** The models' input $x$ is composed of binary 3D multi-plane gene spot representations or stain crops in the unimodal models (VAE1 and 2) or a concatenated combination thereof in the multimodal models (VAE3–5). All inputs are mapped to a shared latent space $z$ via a single encoder. In multimodal models, reconstruction ($x'$) is performed by two separately trained decoders.

Due to differences in data modalities and representations, separate loss functions were required to optimize each output: For stain images, a simple $L_2$ loss function was used. Conversely, as the TC data was represented as a multi-plane binary matrix, formulating the task as a multi-class segmentation problem, a Dice loss function was chosen to handle the class imbalance in terms of the disproportionate foreground and background pixels [36,37]. Dice loss was then computed as twice the intersection of the predicted and ground truth masks divided by the sum of the areas of the two masks (Equation 1). To further mitigate the extreme sparsity challenge, the Dice loss function was combined with Focal loss with the two terms weighted 0.8:0.4. Focal loss is a dynamically scaled cross-entropy function that adjusts pixel contribution based on classification difficulty, amplifying the impact of hard misclassified pixels while suppressing the influence of high-confidence predictions, effectively preventing small errors from dominating the optimization process [38] (Equation 2). Moreover, as empirical evaluation revealed an expected reconstruction bias towards highly expressed genes, genes were stratified based on expression, and a dynamic gene weighting scheme was implemented: a gene's weight was set as the inverse of its expression and initially rescaled to the total TC of the corresponding cell. Weights were calculated for each cell, averaged per class within each batch, and further rescaled to a standardized range. This weighting scheme encouraged the model to reconstruct all genes equally and not neglect less expressed genes in favor of highly expressed ones. In all experiments, the KL divergence term was omitted (β = 0) from the loss function, to prioritize reconstruction fidelity over LS regularization.

$$\text{Equation 1:} \quad Dice\ loss = 1 - \frac{2 \times |X \cap Y|}{|X| + |Y|}$$

$$\text{Equation 2:} \quad Focal\ loss = -\alpha_t(1 - p_t)^\gamma \times log(p_t)$$

## Classical phenotyping

CP, an open-source image analysis software that witnessed widespread use and remarkable success as a standard tool for image-based profiling, was used to conduct classical morphological phenotyping [26,27]. CP features, rooted in classical computer vision algorithms, have been previously shown to contain biologically relevant information regarding cell type and state [39,40]. In order to maximize the extracted feature space, stain crops were input as dual-channel images and processed with measurement modules including MeasureImageQuality, MeasureObjectIntensity, MeasureObjectIntensityDistribution, MeasureTexture, and MeasureGranularity. To evaluate the impact of spatial neighborhood inclusion, small (CPsm, 100 px per side) and large (CPlg, 300 px per side) square crops were processed in the CP pipeline. Feature selection was conducted using the PyCytominer package with the variance_threshold, correlation_threshold, drop_na_columns, and blocklist operations [41].

## Data processing and clustering

LSs were obtained by sampling the trained models' posterior latent distributions, as previously described [42], yielding a single 50-dimensional vector per cell. Using various tools from the scverse Python ecosystem [43], all individual datasets–TC, CP-derived, and LSs–were structured as AnnData objects [44] and handled as modalities in MuData objects [45]. Unsupervised clustering was performed using scverse's implementation of the Leiden algorithm, which improves upon the Louvain algorithm [46] by introducing a refinement step in both the modularity optimization and community aggregation phases, yielding better partitions and providing an explicit guarantee of community-connectedness [47]. While clustering was performed on the full datasets without prior embedding, UMAPs were produced using the Scanpy Python library with a minimum distance of 0.05, and annotated with true labels to enable the visual structural inspection of the dataset embeddings [31].

To account for differences in feature space size and statistical distribution across each dataset as well as the TC gene subsets, Leiden hyperparameters were tuned in an unbiased manner: For individual dataset clustering, the $k$ parameter in the $k$-nearest neighbors ($k$-NN) graph construction and the resolution parameter γ in Leiden's modularity function were optimized via grid search, using silhouette score as an objective function. For multiplex clustering, where layered *k*-NN graphs with common vertex sets but distinct edge sets are first constructed, the *k* and γ parameters were inherited from the corresponding individual dataset cluster optimization, while the modality weight hyperparameter was optimized via grid search, again using the average silhouette score as the objective function. Additionally, a lower cluster count bound of 5 was imposed during all Leiden hyperparameter tuning. Conversely, upstream clustering during VAE model validation and empirical hyperparameter optimization was performed using fixed default parameters (*k* = 15, γ = 1).

Clustering results were evaluated using multiple metrics as implemented by the scikit-learn Python library [48]. These included the internal metric silhouette score as a measure of intracluster cohesion and intercluster separation [49]; and the external metrics homogeneity [50], adjusted mutual information [51], and Jaccard scores [52] as measures of cluster purity, agreement, and overlap in terms of the true labels, respectively. As the Jaccard score is computed on a per cluster-label pair basis, the Hungarian algorithm [53]–as implemented by the SciPy Python library–was used to solve the linear assignment problem and determine the optimal cluster-label correspondence [54].

## Statistical analysis and modeling

To initially assess how well each feature space captured biologically relevant variation, both statistical testing and predictive modeling approaches were adopted. For the former, a permutational multivariate analysis of variance (PERMANOVA) was performed using the true labels as groupings. Using the scikit-learn [48], scikit-bio [55], and SciPy Python libraries [54], the datasets were first standardized before computing a pairwise Euclidean distance matrix and conducting a pairwise PERMANOVA with 999 permutations separately for each true label pair. P-values were adjusted using Benjamini-Hochberg false discovery rate (FDR) correction [56]. Furthermore, considering the imbalance in true label size, a permutational analysis of multivariate dispersion (PERMDISP) was also run to enable appropriate PERMANOVA interpretability.

To complement the statistical analysis, a predictive modeling approach was additionally applied to assess how well each feature space enabled classification of cell types: Using the scikit-learn Python library [48], a random forest classifier with 1,000 trees and balanced class weighting was trained on the standardized feature spaces in a stratified 5-fold cross-validation scheme, ensuring true label distribution remained consistent. In each iteration, the model was trained on four folds and evaluated on the remaining fold, with the test fold rotating across iterations. Balanced accuracy was then computed as the per-class recall for each test fold and averaged across all cross-validation iterations.

The relationship between latent morphological features and gene expression was examined using ridge regression, with LS2 as a multivariate predictor matrix and each gene's TC as the response variable (sci-kit learn Python library, [48]). Cells were randomly split into 80% training and 20% test sets and the model was fit separately for each gene with an L2 regularization parameter α = 1. $R^2$ scores were then computed for the test set to quantify predictive performance. Statistical significance was assessed using the univariate linear regression F-test and Benjamini-Hochberg FDR correction, selecting the lowest adjusted p-value among the LS2 latent variables for each gene.

# RESULTS

## Model validation

Empirical testing showed that the clustering performance of the VAE embeddings was largely insensitive to hyperparameter tuning. Variations in latent vector (z) size or optimization algorithm had no appreciable effect across data modalities. Instead, the performance was heavily influenced by the choice of loss function and the data representation used for the training as well as some modality-specific parameters. This prompted a focused study of the configuration of the transcript-only (VAE1), stain-only (VAE2), and multi-modal models (VAE3–5) to evaluate how their respective design choices impacted the structure and biological relevance of the learned LSs.

Specifically, compared to training on staining crops, the chosen multiclass representation of TC for VAE1 presented a substantially more challenging learning problem that warranted the empirical tuning of spot padding, learning rate, and weighting of loss function terms to accommodate the increased complexity. Experiments showed that spot padding had no significant impact on model performance beyond a minimal threshold, while crop size was empirically selected to retain sufficient transcript signal without incurring unnecessary computational cost (Table 1). In contrast, embedding validation results varied greatly depending on the number of training epochs (Table 2) suggesting that prolonged training led to overfitting that prioritized reconstruction over encoding class-discriminative features.

Experiments with VAE2, trained exclusively on nuclear and membrane stain crops, identified crop size as the critical parameter affecting model performance: Smaller crops often excluded portions of the cells and markedly reduced cluster homogeneity score, likely due to omitting key morphological information necessary for meaningful representation learning. In contrast, larger crops incorporated additional spatial context from neighboring cells leading to improvements in clustering performance (Table 3). Training the multi-modal models (VAE3–5), which jointly encoded TC and stain images, introduced additional computational challenges owing to the increased input dimensionality from combining the stain crops with the multi-plane representation of TC data. To alleviate the computational burden, a crop size of 150 px (VAE3) or 200 px (VAE4–5) was adopted, and genes with fewer than 100 detected transcripts across the tissue were excluded, resulting in a more compact TC representation composed of 196 genes (Table 1). Additionally, a weight decay factor of 0.0004 was applied as a form of L2 regularization to mitigate overfitting and contribute to training stability. Weight decay penalizes large parameter values by adding a regularization term to the loss function, thereby encouraging smaller weights in the variational autoencoder (VAE). This not only reduces the risk of overfitting but can also help limit the magnitude of gradients during training, indirectly reducing the likelihood of exploding gradients [57].

## Transcript count and stain latent spaces captured biologically relevant information

The GT-annotated spatial map showed a highly organized tissue structure, with several cell types forming distinct layers (Figure 2A and B). Notably, smooth muscle cells, Paneth cells, and enterocytes exhibit clear spatial arrangement, with enterocytes following a functional maturation gradient along the villus axis [58]. Collectively with the availability of morphological stains, this dataset captures key aspects of biological variability and organization, offering a relevant context to evaluate approaches for multi-modal integration. Of note, the true label distribution displayed a strong disparity in cell type abundance, spanning over a 40-fold range.

GT-annotated UMAPs enabled the preliminary qualitative assessment of the degree to which the different feature spaces encoded variation across true labels (Figure 3): the TC embedding showed strong, well-defined structure, as expected. Interestingly, several of the LSs exhibited low to moderate organization, despite their compressed representation and limited training. In contrast, the CPsm and CPlg UMAPs displayed no discernible pattern. This was quantitatively evaluated through statistical testing via PERMANOVA (Figure S1) and through predictive modeling via cross-validated classification performance (Figure 2C). PERMANOVA revealed that TC and LS3–5 captured the strongest separation between true labels, with all or nearly all pairwise comparisons yielding significant differences. While LS1 and LS2 showed more non-significant comparisons—primarily among rare or underrepresented cell types like tuft cells, myenteric plexus cells, interstitial cells of Cajal, and circulating follicular B cells—their ability to distinguish more common types remained highly promising. CP-derived features performed substantially worse, though CPlg showed some signal in smooth muscle cells, suggesting a certain degree of specificity even in the absence of global structure.

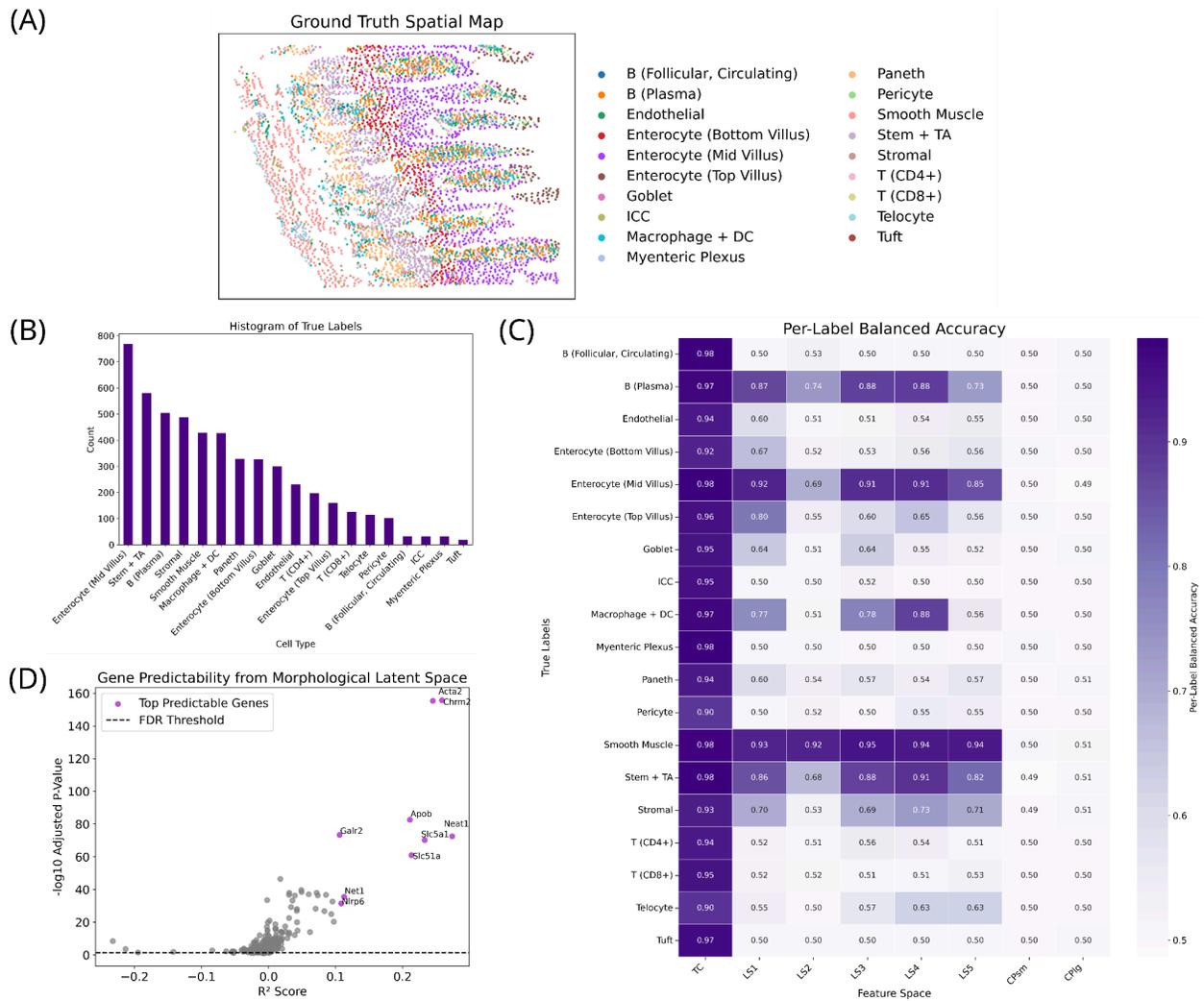

**FIGURE 2 | Overview of dataset characteristics and predictive performance.** (A) Spatial map of the tissue sample annotated by true labels (ICC: interstitial cells of Cajal, DC: dendritic cells, TA: transit-amplifying cells). (B) Histogram showing the distribution of true labels. (C) Random forest 5-fold cross-validation results showing per-label balanced accuracy across all feature spaces: transcript count (TC), VAE latent spaces (LSs), and CellProfiler features from small (CPsm) and large (CPlg) crops. (D) Performance of ridge regression models predicting gene expression from the morphological latent space (LS2) representing stain images. P-values were corrected for multiple testing using the Benjamini–Hochberg false discovery rate (FDR) method with α = 0.05.

Although distance-based statistical analyses are non-parametric and generally robust, PERMANOVA carries implicit assumptions about mean-variance relationships, where significance may arise from either centroid shifts or differences in within-group dispersion [59]. To disentangle these effects, we applied PERMDISP [60], a multivariate extension of Levene's homoscedasticity test, as previously described [61]: a non-significant PERMDISP result supports attributing PERMANOVA significance to a location effect, whereas a significant PERMDISP result suggests that dispersion differences contributed to the observed effect. PERMDISP results indicated that TC, LS1, and LS3 had a similar number of significant dispersion differences, suggesting that dispersion differences contributed to the observed structure, while LS2, LS4, and LS5 exhibited fewer, suggesting differences were more likely driven by centroid shifts. Complementing this, classification results offered further insight into how well each space preserved biologically meaningful information (Figure 2C): Balanced accuracy for TC was consistently high across all cell types (>0.90), indicating strong predictive power. Encouragingly, LS1 and LS3–5 achieved moderate performance in several cell types including smooth muscle cells, mid-villus enterocytes, stem and transient amplifying cells, plasma B cells, and macrophages and dendritic cells. Even LS2, exclusively representing stain images, captured smooth muscle cell identity with high balanced accuracy. CP-derived features performed the poorest with balanced accuracy scores reflecting random chance.

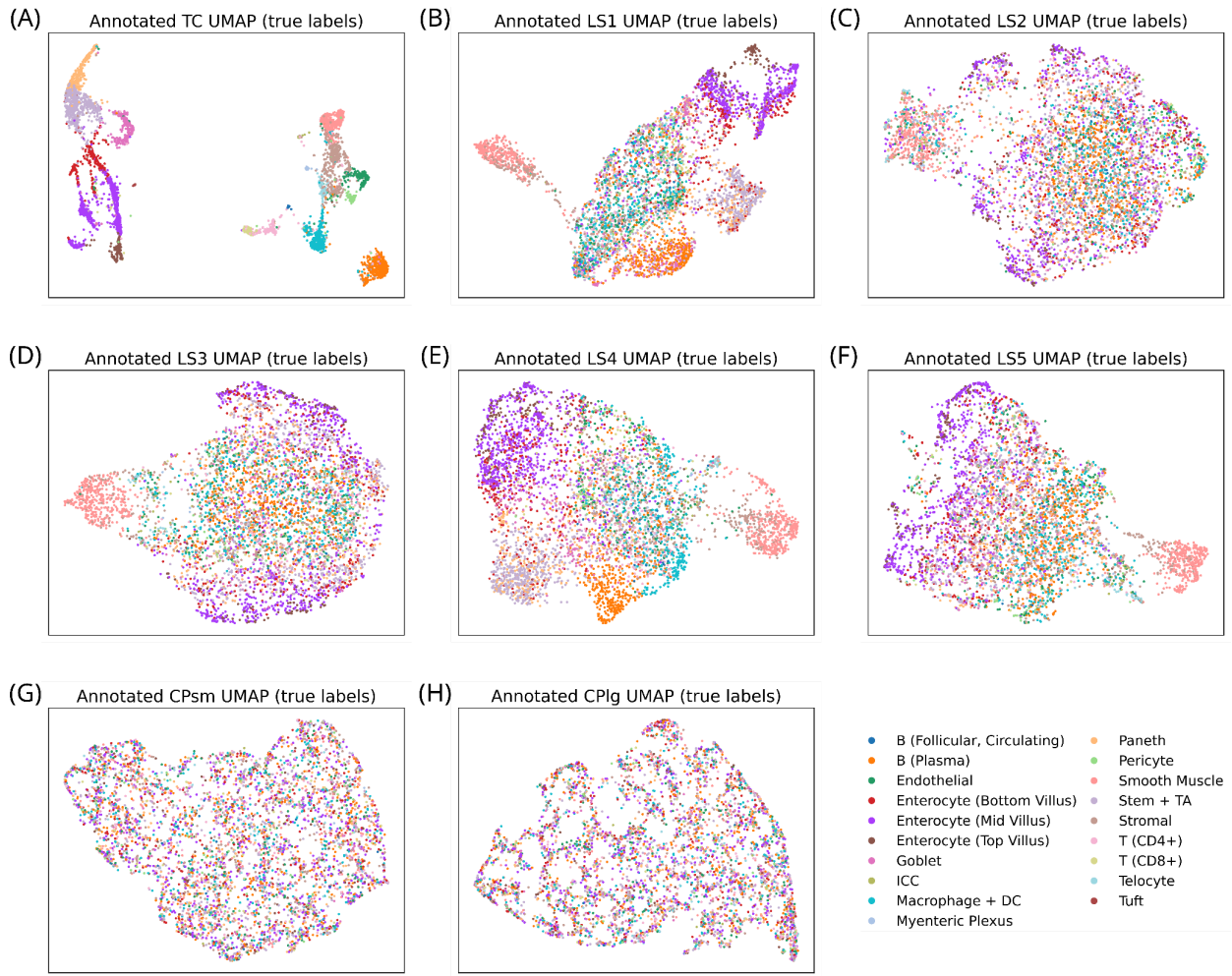

**FIGURE 3 | True label organization across UMAP projections of feature spaces.** (A) UMAP of the

transcript count feature space (TC). (B–F) UMAPs of the VAE latent spaces (LSs). (G–H) UMAPs of the CellProfiler feature spaces from small (CPsm) and large (CPlg) crops.

Unsupervised Leiden clustering largely echoed these patterns. While TC outperformed all other feature spaces across clustering metrics (Figures S2 and S3), LS4 and LS1 in particular showed the strongest clustering performance among the LSs, followed by LS3, LS5, and LS2. Notably, CPsm and CPlg showed slightly higher silhouette scores than TC, though this was not reflected in other metrics, which were substantially lower. Jaccard-based cluster-label matching revealed both strengths and limitations: Even TC had unmatched clusters for four true labels, and LSs had more unmatched cases, yet importantly, in matched labels, LSs occasionally outperformed TC. Jaccard scores of LS clusters surpassed those of TC for smooth muscle cells (LS1, LS3, LS4), mid-villus enterocytes (LS1, LS4), stromal cells (LS1, LS3), and CD8+ T-cells (LS2), indicating that useful biological signal can indeed emerge from these compressed representations.

To examine the relationship between morphological features captured by LS2 and gene expression, a ridge regression was applied to predict TC from LS2. As depicted in Figure 2D, LS2 demonstrated modest ability to predict TC with six genes exceeding an $R^2$ score of 0.2 (*Neat1*, *Chrm2*, *Acta2*, *Slc5a1*, *Slc51a*, and *Apob*) and three additional genes exceeding an $R^2$ of 0.1 (*Net1*, *Nlrp6*, and *Galr2*). However, global predictive power remained limited with near-zero or negative $R^2$ scores for the majority of the remaining genes.

## Joint clustering of transcript counts and latent spaces enhanced cluster homogeneity

To evaluate how spatial context might complement standard count-based gene expression analysis, the TC feature space was jointly clustered with each LS or with CP-derived features, incorporating them as separate layers in the *k*-NN graph constructed for Leiden clustering. With this integrative approach, pairing TC with each of LS2–5 led to clear improvements in homogeneity scores—indicating greater within-cluster purity—even with fewer clusters formed (6–8 vs. 16) (Figures 4 and S4). Moreover, in terms of cluster-label Jaccard matching, integrating TC with the LSs improved Jaccard scores for mid-villus enterocytes and smooth muscle cells, though it hindered overall true label recovery. Notably, this approach also uncovered structure that was not captured by TC alone: a cluster was matched to stromal cells in TC+LS2 but not in TC alone. Conversely, joint clustering of TC with CP-derived features resulted in complete collapse in performance.

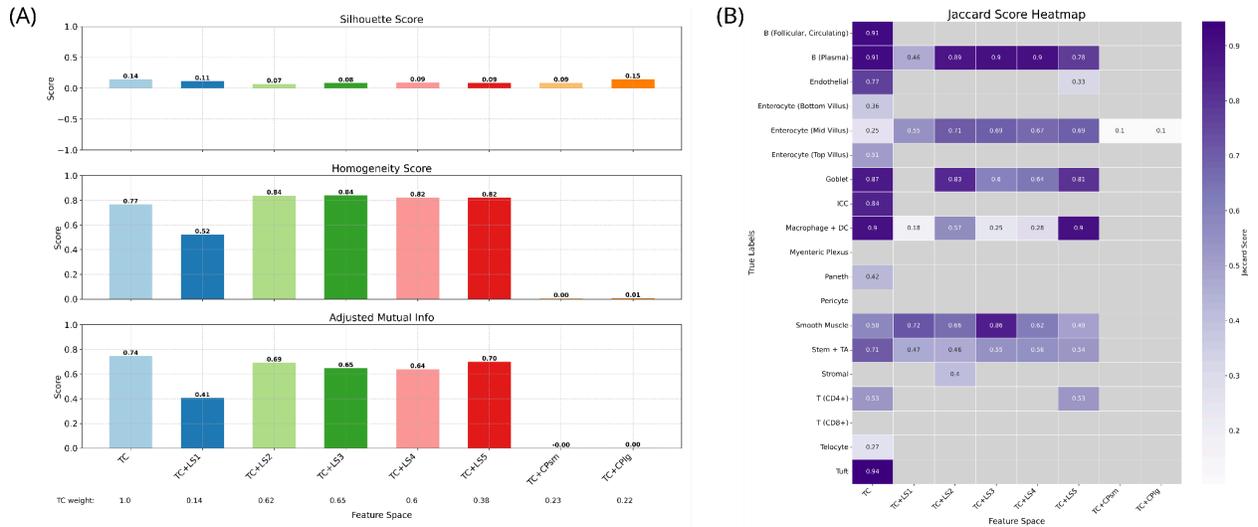

**FIGURE 4 | Clustering metrics across multiplexed feature spaces.** (A) Mean silhouette score (top) of the transcript count (TC) feature space combined with each latent space (LS) or CellProfiler feature set from small (CPsm) or large (CPlg) crops, and homogeneity (middle) and adjusted mutual information scores (bottom) with respect to the ground truth partition. The TC modality weight selected by optimizing multiplex Leiden clustering hyperparameters (objective: silhouette score) is shown below each corresponding feature space pair. (B) Jaccard scores for clusters matched to true labels via the Hungarian algorithm across the multiplexed feature spaces.

Finally, to explore the benefit of joint clustering of TC with morphological features under conditions of limited gene expression data, this constraint was simulated using three gene subsetting approaches: a PCA-based method (25 genes), a GT-based method (19 genes), and a random selection approach (19 genes). Despite the reduced gene set, overall clustering trends remained unchanged (Figure 5 and S5). While silhouette and adjusted mutual information scores remained similar to TC-only clustering, homogeneity scores consistently improved with the inclusion of LS2 across all gene selection approaches. Of note, as part of the multiplex Leiden clustering, we optimized the weight of incorporation of each feature space, and the resulting weights—3%, 11%, and 5% for the GT-based, PCA-based, and random subsets, respectively— suggest that small contributions from morphological features can enhance cluster purity under these conditions.

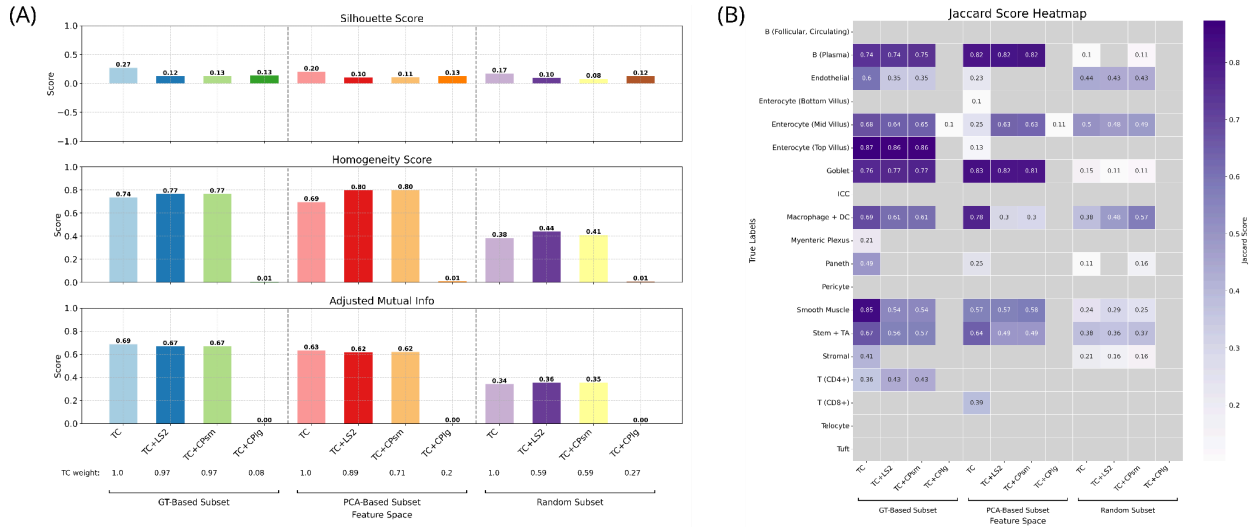

**FIGURE 5 | Clustering metrics across multiplexed feature spaces using gene subsets.** (A) Mean silhouette score (top) of the transcript count (TC) feature space combined with each latent space (LS) or CellProfiler feature set from small (CPsm) or large (CPlg) crops, and homogeneity (middle) and adjusted mutual information scores (bottom) with respect to the ground truth partition. Gene subsets were defined using three strategies: true label-based ranking (GT-based, left), PCA loading-based ranking (middle), and random gene selection (right). The TC modality weight selected by optimizing multiplex Leiden clustering hyperparameters (objective: mean silhouette score) is shown below each corresponding feature space pair. (B) Jaccard scores for clusters matched to true labels via the Hungarian algorithm across the multiplexed feature spaces.

# DISCUSSION

While DL has driven major advances in multi-modal integration of ST with complementary data types, most studies to date have focused on integration with histological stains [9–13], chromatin images [14], or scRNA-seq [15–18]. In contrast, the integration of ST with paired nuclear and membrane stain images remains relatively underexplored, despite their routine availability in experimental workflows and potential to inform on spatial context and cellular morphology. This study aimed to address this gap by applying VAEs to learn a low-dimensional latent representation of TC from a publicly available murine ileal tissue MERFISH dataset—a single image with 241 genes and 5192 cells—while incorporating or excluding corresponding nuclear and membrane stain images [24]. In addition to its relatively small size, this dataset contains starkly imbalanced cell types, and its accompanying membrane stain delineates $Na^+/K^+$-ATPase protein expression, which is strictly basolaterally polarized in epithelial cells [62], thereby recapitulating real-world experimental constraints. Furthermore, robustness and interpretability were prioritized over optimizing for peak model performance to enable the evaluation of the fundamental utility of VAEs in this context: this was achieved by using bounding-box crops in lieu of precise cell segmentation, employing a standard VAE architecture without attention mechanisms or clustering-aware training (e.g. a gaussian mixture model prior), omitting systematic model hyperparameter tuning, and sampling latent variable distributions instead of using mean embeddings for downstream analyses. This approach aligns with prior findings demonstrating that, in smaller dataset regimes, complex DL models may not yield superior performance over simpler architectures [63], especially in the context of VAEs which retain the ability to learn robust disentangled feature representations [64]. However, unlike previous studies that relied on aggregated gene count vectors to learn a latent representation of ST data [14,32,33], TC data were represented as multi-plane binary matrices for model input to preserve local spatial context.

Two integration strategies were examined: one involved a joint VAE LS with a dual-decoder architecture for simultaneous learning and subsequent LS clustering, while the other employed multiplex Leiden clustering of TC data together with LSs extracted from independently trained VAEs. These VAEs were trained on TC, nuclear and membrane stain images, or a combination thereof. Additionally, classical morphological features were extracted using CP to provide a reference for evaluating the VAE-derived morphological embeddings. This proof-of-concept study revealed that VAE-based morphological embeddings captured biologically relevant variability and improved cluster homogeneity when integrated with TC via the multiplex Leiden algorithm, and demonstrated that such embeddings offer significant advantages over classical phenotyping approaches.

An initial assessment of the extent to which the TC dataset, VAE-derived LSs, and CP-derived features encoded biologically meaningful variation was conducted via both statistical testing and predictive modeling approaches. As VAE3–5 were trained on multi-modal input, though constrained to a 76%-gene subset to curb model complexity, LS3, LS4, and LS5 aligned best with TC by statistically separating all or nearly all labels. This is in contrast to LS1 and LS2, trained on single modalities, as their performance appeared to be limited by label rarity underscoring the well-established bias of VAEs towards dominant features in imbalanced inputs [65]. However, an analysis of dispersion differences with a pairwise PERMDISP test revealed that for TC, LS1, and LS3, a substantial number of significant dispersion differences were observed, suggesting that some of the significant PERMANOVA comparisons may have been influenced by differences in within-group variability rather than true centroid shifts [59]. Although this could reflect a class imbalance artifact, it may also indicate genuine biological heterogeneity in cell types as dispersion heterogeneity has been recognized as a meaningful metric across multiple biological scales, from gene expression [66,67], to organelle abundance [68], to species diversity [69]. Conversely, LS2, LS4, and LS5 exhibited lower differences in dispersion, lending credence to a primarily centroid shift-driven separation [59]. While this indicates that LSs captured biologically-relevant variation, statistically significant separations do not necessarily imply that the data is structured in a manner permissive of accurate classification without prior knowledge of the true labels. Indeed, cross-validated classification performance (balanced accuracy) showed that fewer cell types were accurately recalled using the LSs, especially LS2—an expected, albeit still remarkable, outcome given that VAE2 was trained on stain images only. In contrast, CP-derived features unambiguously underperformed across both the statistical and classification approaches, highlighting the superior capacity of VAEs in extracting biologically-relevant morphological features over classical computer vision methods—at least under the constraint of bounding-box as opposed to precise cell border segmentation.

To evaluate the practical utility of these feature spaces without explicit reliance on prior knowledge, we applied unsupervised Leiden clustering [47]. While this approach captured fewer cell types than classification, it revealed consistent patterns and highlighted opportunities for improvement. Notably, we observed that LS5, representing TC and both stains, was outperformed by LS1 (TC alone) and LS4 (TC + membrane stain) in terms of homogeneity and adjusted mutual information scores despite representing more information. This shortfall may stem from the well-documented modality underutilization or collapse problem in multi-modal VAEs whereby models trained on small datasets tend to over-rely on dominant channels, discarding complementary information and degrading LS quality [70,71]. This suggests that a larger dataset or a VAE architecture that explicitly addresses modality underutilization may be requisite for realizing the added value of the three-input setting. Building on these findings, we then explored how TC might be complemented by additional feature spaces with multiplex Leiden clustering. In addition to LS2 and CP-derived features, we also jointly clustered TC with LSs from VAE1 and VAE3–5, as the TC component of their input retained local spatial context that could provide complementary information. Interestingly, this multiplex clustering of TC with LSs but not with CP-derived features, boosted homogeneity scores, improved the Jaccard scores for certain labels (namely mid-villus enterocytes and smooth muscle cells), and enabled the matching of stromal cells (TC+LS2), though these improvements were accompanied by trade-offs including a lower adjusted mutual information score and a decline in overall label recovery. As homogeneity score is not agnostic to class imbalance, the boost in this metric can be largely explained by the improved identification of the predominant mid-villus enterocyte population, overshadowing the weaker performance on less abundant cell types and the decrease in cluster count relative to TC alone [72]. Concordant findings were observed when jointly clustering a subset of the TC feature space with LS2, where the homogeneity score improvement was largest in the PCA-based gene subset—coinciding with the greatest increase in mid-villus enterocyte Jaccard score.

The observed increase in homogeneity score upon the multiplex clustering of TC with LS2–5 and the improved recovery of select labels indicate that the VAEs encoded biologically meaningful variation that holds promise, even if it did not translate into overall clustering improvements under the constraints imposed herein. This is supported by the ridge regression analysis that demonstrated the predictability of a subset of genes from LS2—a latent representation of solely nuclear and membrane stain images. These included transcripts highly enriched in smooth muscle cells (*Acta2*, *Chrm2,* and *Galr2*), explaining the label's high Jaccard score in LS2 clustering, which in turn is likely underlain by the morphological distinctiveness of smooth muscle cells and their spatial co-enrichment in the muscularis propria tissue layer. Moreover, *Acta2*, is also enriched in stromal cells, which were not recovered in TC clustering but rather only upon its multiplex clustering with LS2. As *Acta2* encodes α-actin 2, a cytoskeletal protein with clear morphological relevance [73], its predictability from LS2 likely reflects VAE2's ability to capture subtle morphological cues such as differences in cell shape or nuclear organization that correlate with *Acta2*-expressing cells. Conversely, *Neat1*, the most strongly predictable gene from LS2, encodes the long non-coding RNA (lncRNA) nuclear enriched abundant transcript 1, an integral structural component of nuclear paraspeckles [74], suggesting that VAE2 learned its expression pattern in a more direct manner based on nuclear morphology. This reinforces the notion that latent representations retain interpretable biological signals.

Collectively, these findings highlight both the limitations and potential of VAEs for augmenting ST data with common, easily acquired stains. While the current approach was constrained by dataset size and segmentation limitations, several refinements could enhance its applicability without deviating from real-world constraints. Simple methodological improvements, such as systematic hyperparameter tuning, the leveraging of both the mean and variance vectors of the LS instead of sampling the latent variable distributions, or employing alternative clustering techniques like Gaussian mixture models, sub-clustering, or consensus clustering, could yield more robust insights. Notably, modularity-based community detection algorithms, such as Leiden clustering, are known to suffer from a resolution limit that biases against rare cell types, making them suboptimal for imbalanced datasets [75]. Beyond these refinements, more sophisticated strategies warrant exploration. β-VAEs could improve the disentanglement of biologically relevant features [76,77], contrastive learning or class-aware loss functions could enhance feature separation, and pretraining on larger ST or tissue imaging datasets could improve representation learning while keeping the approach accessible and broadly applicable.

## CODE AND DATA AVAILABILITY

The code used for the VAE training and data analysis, the model inferences, and the CP pipeline is available at https://github.com/ciminilab/2025_Moser_Hamid_submitted

## COMPETING INTERESTS

B.A.C. serves as a scientific advisor for companies that use image-based profiling (Medici Therapeutics). All other authors declare no competing interests.

## ACKNOWLEDGEMENTS

This work was funded by NIH grants P01AI148102 and P41GM135019 as well as grant 2023-329649 from the Chan Zuckerberg Initiative DAF, an advised fund of Silicon Valley Community Foundation. The funders had no role in study design, data collection and analysis, decision to publish, or preparation of the manuscript. The authors greatly thank other members of the Broad Institute Imaging Platform for helpful discussions.

# TABLES

**TABLE 1** Model configurations for VAE1–5. Each column lists the input modalities, crop sizes, and training hyperparameters used for the respective model. TC refers to transcript count data represented as multi-plane binary images. For multimodal models (VAE3–5), transcript and stain inputs are concatenated before encoding. Spot padding refers to the size of the convolutional kernel used to expand each transcript spot into a local binary patch where a padding of $p$ produces a $(p + 2) \times (p + 2)$ kernel.

|  | VAE1 | VAE2 | VAE3 | VAE4 | VAE5 |
|---|---|---|---|---|---|
| **Input** | TC<br>*Crop: 100 px*<br>*Genes: 241* | DAPI<br>*Crop: 300 px* | TC<br>*Crop: 150 px*<br>*Genes: 182* | TC<br>*Crop: 200 px*<br>*Genes: 182* | TC<br>*Crop: 200 px*<br>*Genes: 182* |
|  |  | Membrane stain<br>*Crop: 300 px* | DAPI<br>*Crop: 150 px* | Membrane stain<br>*Crop: 200 px* | DAPI<br>*Crop: 200 px* |
|  |  |  |  |  | Membrane stain<br>*Crop: 200 px* |
| **Spot padding** | 3 | N/A | 4 | 4 | 4 |
| **Learning rate** | 5e-4 | 5e-5 | 5e-4 | 5e-4 | 5e-4 |
| **Batch size** | 8 | 16 | 16 | 16 | 20 |
| **Epochs** | 15 | 500 | 20 | 30 | 30 |
| **Adam's ε** | 1e-8 | 1e-8 | 1e-6 | 1e-6 | 1e-6 |
| **Weight decay** | N/A | N/A | N/A | N/A | 1e-5 |

**TABLE 2** Cluster homogeneity scores and proportion of label-matched clusters for VAE embeddings trained on TC across increasing training epochs.

| Epochs | Homogeneity score | Cluster correlations |
|---|---|---|
| **100** | 0.342 | 9/7 |
| **50** | 0.275 | 9/12 |
| **20** | 0.337 | 11/13 |
| **15** | 0.416 | 11/11 |

**TABLE 3** Cluster homogeneity scores and proportion of label-matched clusters for VAE embeddings trained on stain images across increasing crop sizes.

| Crop Size | Homogeneity score | Cluster correlations |
|---|---|---|
| **100** | 0.055 | 4/10 |
| **150** | 0.091 | 3/13 |
| **200** | 0.129 | 3/11 |
| **250** | 0.147 | 4/12 |
| **300** | 0.158 | 4/11 |

# Supplementary Information

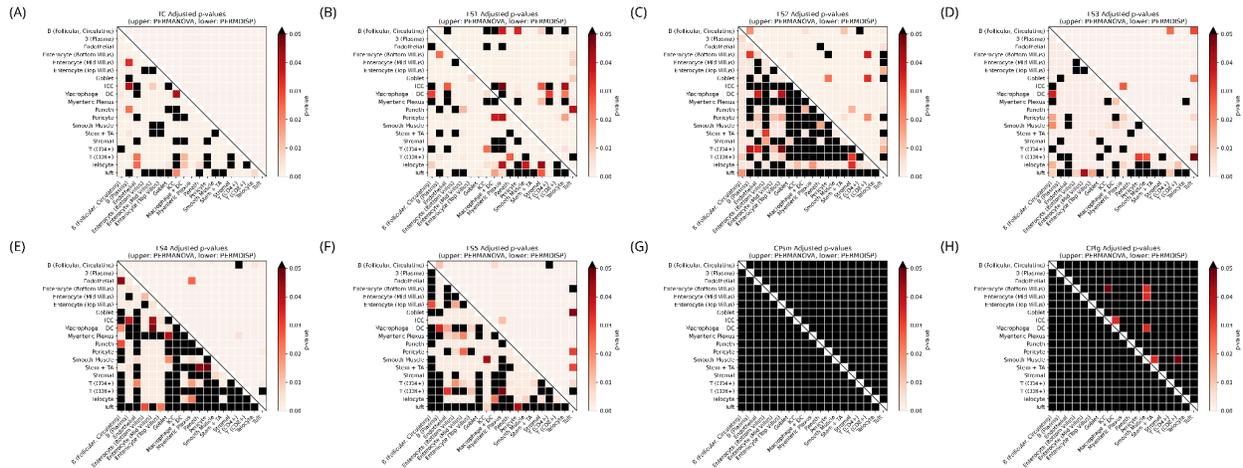

**FIGURE S1 | Pairwise statistical comparisons between true labels in each feature space.** (A–H) Half-matrix heatmaps showing pairwise permutational multivariate analysis of variance (PERMANOVA, upper triangle) and permutational analysis of multivariate dispersion (PERMDISP, lower triangle) results for each feature space: transcript count (TC), VAE latent spaces (LSs), and CellProfiler features from small (CPsm) and large (CPlg) crops. Multiple testing correction was applied using the Benjamini-Hochberg false discovery rate (FDR) method; p-values are color-mapped, with values above 0.05 shown in black.

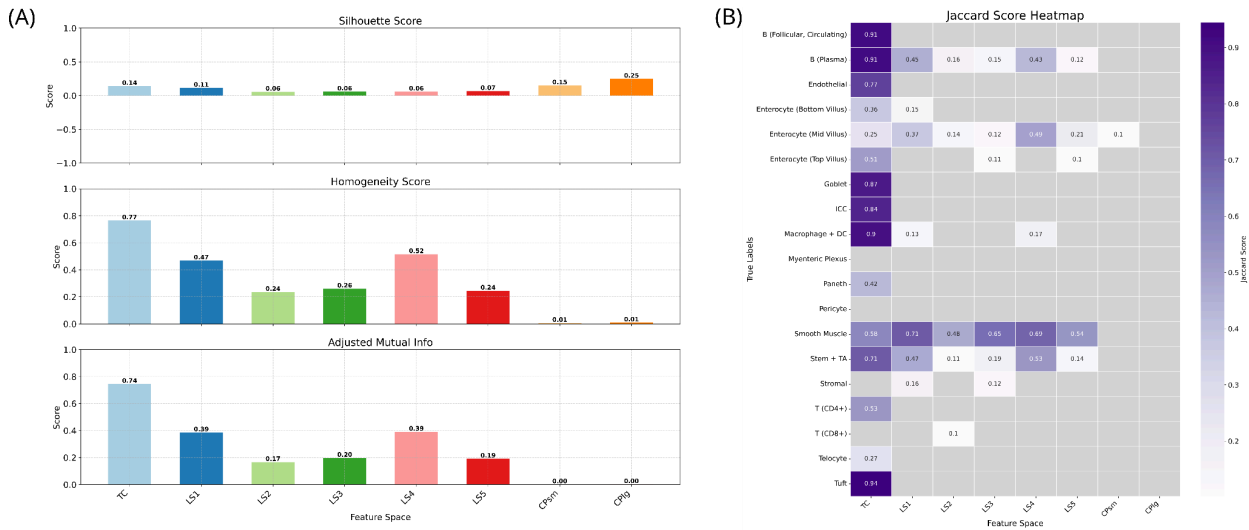

**FIGURE S2 | Clustering metrics across the feature spaces.** (A) Silhouette score (top), and homogeneity (middle) and adjusted mutual information scores (bottom) with respect to the ground truth partition. (B) Jaccard scores for clusters matched to true labels via the Hungarian algorithm across the feature spaces: transcript count (TC), VAE latent spaces (LSs), and CellProfiler features from small (CPsm) and large (CPlg) crops.

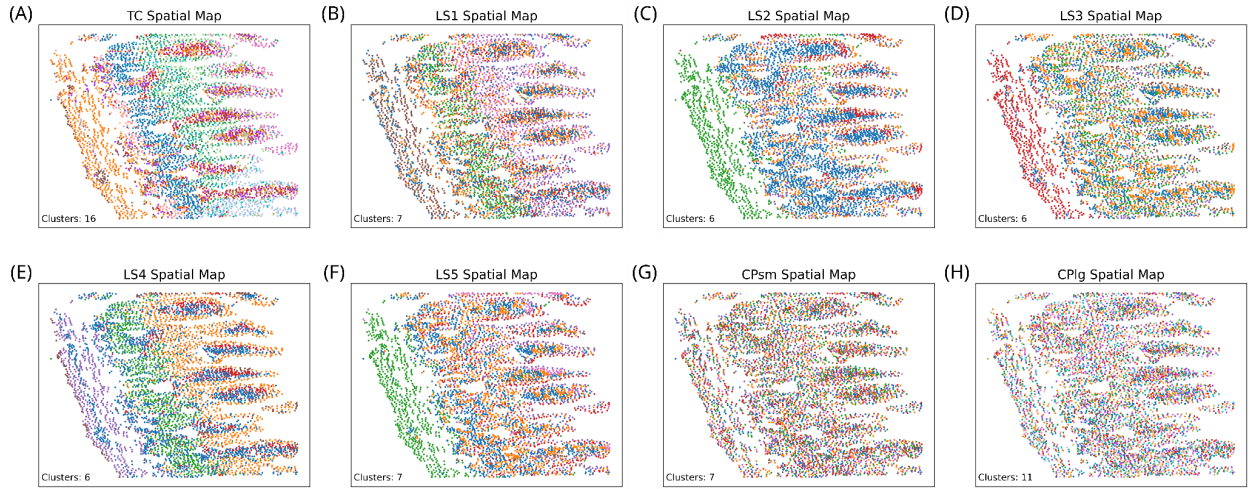

**FIGURE S3 | Spatial organization of clustering results across feature spaces.** (A–H) Spatial map of the tissue sample annotated by Leiden clustering results for individual feature spaces: transcript count (TC), VAE latent spaces (LSs), and CellProfiler features from small (CPsm) and large (CPlg) crops. The number of clusters is indicated in the bottom left of each panel.

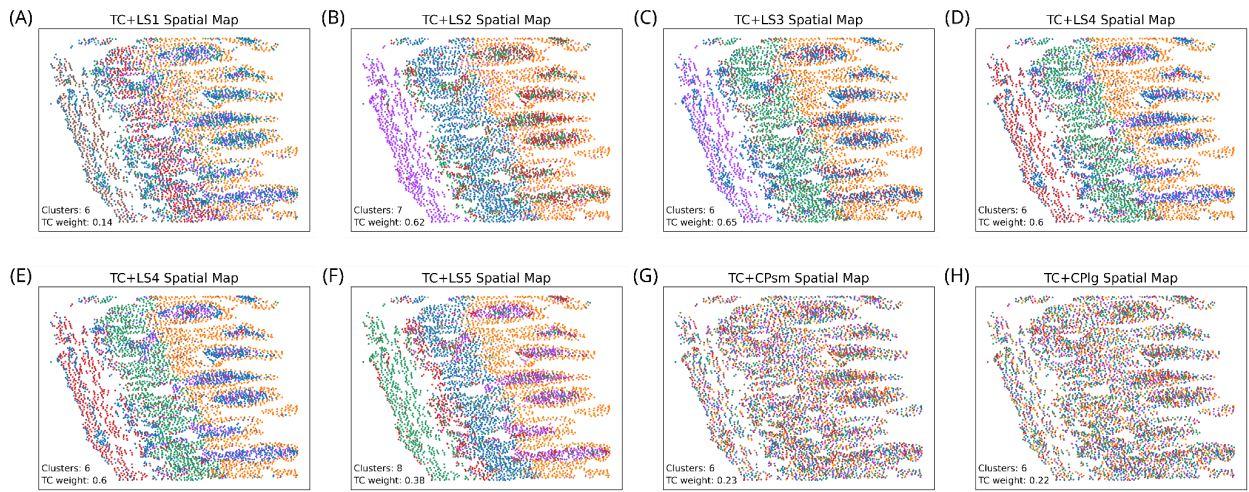

**FIGURE S4 | Spatial organization of clustering results across multiplexed feature spaces.** (A–H) Spatial map of the tissue sample annotated by Leiden clustering results for the transcript count (TC) feature space multiplexed with each VAE latent space (LS) or CellProfiler feature set from small (CPsm) and large (CPlg) crops. The number of clusters and the optimized TC modality weight (based on mean silhouette score) are indicated in each panel.

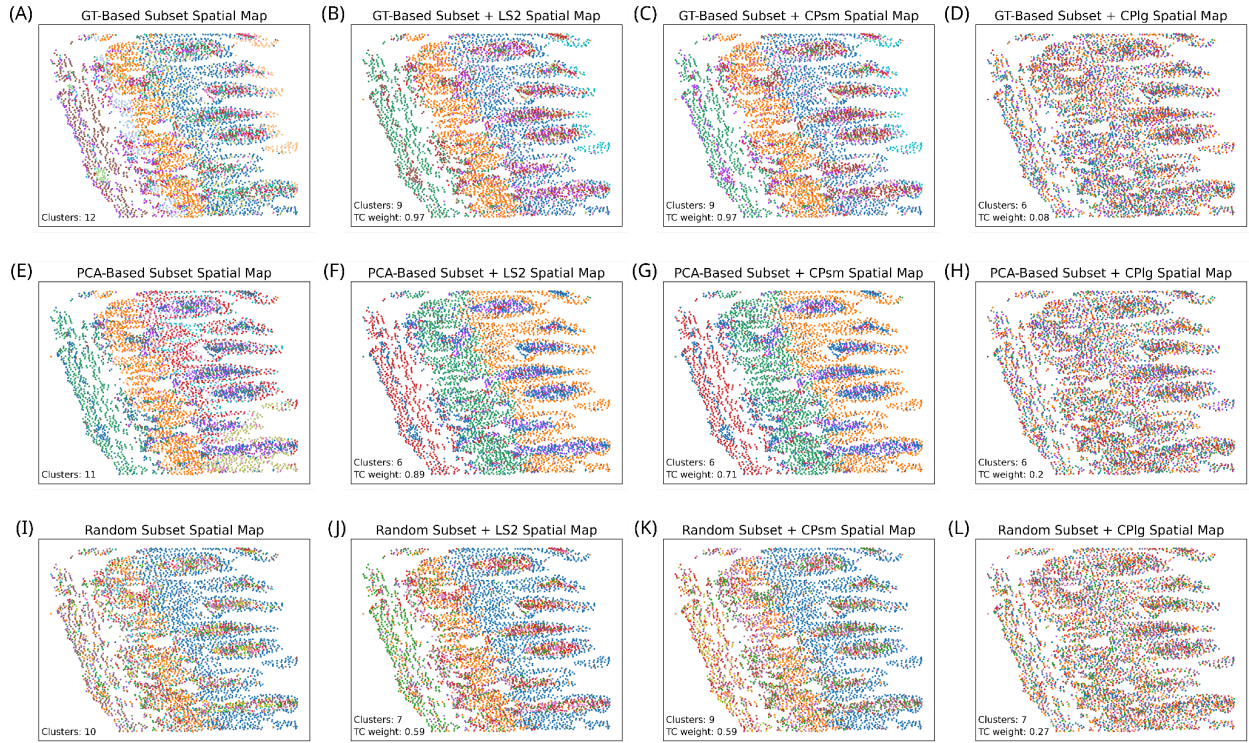

**FIGURE S5 | Spatial organization of clustering results across multiplexed feature spaces using gene subsets.** Spatial maps of the tissue sample annotated by multiplex Leiden clustering results. Rows correspond to different gene subset selection methods: (A–D) ground truth (GT)-based selection, (E–H) PCA-based selection, and (I–L) random selection. Each panel shows clustering based on the transcript count (TC) feature space multiplexed with the morphological latent space (LS2) or CellProfiler features from small (CPsm) or large (CPlg) crops. The number of clusters and the optimized TC modality weight (based on mean silhouette score) are indicated in each panel.